\documentclass[prl,aps,twocolumn,showpacs,superscriptaddress,amsmath,amssymb]{revtex4-1}
\usepackage{graphicx}
\usepackage{epstopdf}
\usepackage{dcolumn}
\usepackage{bm}

\begin{document}

\title{Anisotropic critical magnetic fluctuations in the ferromagnetic superconductor UCoGe}

\author{C. Stock}
\affiliation{NIST Center for Neutron Research, 100 Bureau Drive, Gaithersburg, Maryland 20899, USA}
\affiliation{Indiana University, 2401 Milo B. Sampson Lane, Bloomington, Indiana 47404, USA}

\author{D.A. Sokolov}
\affiliation{School of Physics and CSEC, University of Edinburgh, Edinburgh EH9 3JZ, UK}

\author{P. Bourges}
\affiliation{Laboratoire LŽon Brillouin (UMR12 CEA-CNRS), 91191 Gif-sur-Yvette Cedex, France}

\author{P. H. Tobash}
\author{K. Gofryk}
\author{F. Ronning}
\author{E.D. Bauer}
\affiliation{Los Alamos National Laboratory, Los Alamos, New Mexico 87545, USA}

\author{K. C. Rule}
\affiliation{Helmholtz Zentrum Berlin, D-14109, Berlin, Germany}

\author{A.D. Huxley}
\affiliation{School of Physics and CSEC, University of Edinburgh, Edinburgh EH9 3JZ, UK}

\date{\today}

\begin{abstract}
We report neutron scattering measurements of critical magnetic excitations in the weakly ferromagnetic superconductor UCoGe. The strong non-Landau damping of the excitations we observe, although unusual has been found in another related ferromagnet, UGe$_{2}$ at zero pressure. However, we also find there is a significant anisotropy of the magnetic correlation length in UCoGe that contrasts with an almost isotropic length for UGe$_2$. The values of the magnetic correlation length and damping are found to be compatible with superconductivity on small Fermi surface pockets. The anisotropy may be important to explain why UCoGe is a superconductor at zero pressure while UGe$_2$ is not.
\end{abstract}

\pacs{PACS numbers: 75.40.Gb, 74.70.Tx, 75.50.Cc}
\vspace{-0.8cm}

\maketitle

Experimental evidence pinning down mechanisms responsible for superconductivity (SC) in non-conventional, including high temperature superconductors, is required to build an understanding of these interesting phenomena. In this letter we focus on a family of exotic superconductors in which superconductivity co-exists with ferromagnetism. We report measurements of the microscopic fluctuations of the magnetisation that provide insight into how such fluctuations may give rise to superconductivity and an impetus focussing attention on the role of small Fermi-surface sheets in this mechanism. 

The theoretical idea that longitudinal magnetic fluctuations can mediate odd-parity pairing of equal-spin fermionic electronic excitations giving SC is well established \cite{fay80}. Since this SC can survive in the strong exchange field, present in a ferromagnet, this mechanism is a natural starting point from which to explain SC in three ferromagnets UGe$_2$ (under pressure) \cite{sax00,hux01}, URhGe \cite{aok01,lev05} and UCoGe \cite{huy07}. The rarity of SC in ferromagnets, however, suggests that other special circumstances may additionally be required. 

In all three materials superconductivity is correlated with crossing magnetic transitions, but the magnetic transitions differ. For UGe$_2$ only the magnitude of the moment changes  \cite{hux01}, whereas in URhGe the direction of the moment changes with only a modest change of magnitude \cite{lev05}.
In UCoGe the superconducting temperature ($T_s$) is maximum close to the pressure at which FM is suppressed. SC also exists in the paramagnetic state achieved when ferromagnetism FM is suppressed with either pressure or by changes of composition \cite{slo09}, whereas in the other materials SC has only been observed at temperatures much less than the Curie temperature T$_\text{Curie}$ \cite{har05a}.
Thus only UCoGe provides access to superconductivity at a zero-field quantum critical point separating FM from paramagnetism corresponding to the situation most examined theoretically \cite{rou01}.

Neutron scattering studies of critical magnetic fluctuations in f-electron ferromagnets to date are limited to UGe$_2$ in zero pressure where there is no SC \cite{hux03a,ray04}. Over-damped magnetic excitations were found above and below $T_\text{Curie}$ with an integrated spectral weight that accounts for the measured uniform susceptibility. The q-dependence of the energy spectrum is described by a relaxation process that contrary to pure Landau damping does not vanish as the wavevector $q \rightarrow 0$. 
A detailed microscopic theory is lacking, but non-Landau damping may be brought about from a combination of a large spin-orbit interaction and either multiple bands or impurity scattering. For UGe$_2$ the suppression of the extra relaxation process at low temperature below $T_\text{Curie}$ and the high purity of the samples studied suggest that an impurity-based explanation is unlikely. This is supported by the observation that the magnetic susceptibility is not strongly sample dependent or modified by annealing. Since the extra damping boosts the energy scale of low-q excitations responsible for electron-pairing it is expected to promote magnetically mediated SC.  In this Letter, we report the first detailed INS study of UCoGe in which ferromagnetism and superconductivity occur at atmospheric pressure and identify additional features in the magnetic excitation spectrum that
may bring about magnetically mediated SC close to a magnetic quantum critical point.

Unlike for URhGe and UGe$_2$, FM in UCoGe appears to show some sensitivity to sample quality, with the highest T$_\text{Curie}$ of 3 K found in samples having the highest residual resistance ratio (RRR), although SC is surprisingly robust. All the main features of the phase diagram including an unusual temperature dependence of the critical field for superconductivity are present in samples with modest RRR  and reduced $T_\text{Curie}$ \cite{stev11}. The highest RRR single crystals of UCoGe grown to date \cite{huy09} are too small (mm$^3$) for inelastic neutron scattering measurements. The samples used in this work comprise three large (5, 7 and 14 g) single crystals grown by the Czochralski technique followed by an annealing procedure similar to Ref. \onlinecite{huy09}.  A small piece from the 5 g sample was characterised using magnetisation and heat capacity and found to display bulk FM below 1 K and a T$_{S}$ with a midpoint of 0.38 K.  The normal state RRR of this sample was 4.  Unpolarised diffraction measurements on the 7 g single crystal at 0.3 K with FLEX (HMI) found ferromagnetic Bragg peaks with an estimated ordered moment of $\sim$ 0.05$\mu_{B}$.  This indicates that the large neutron samples were both ferromagnetic and superconducting. Our study focusses on temperatures centred around 20 K where magnetic scattering is peaked owing the need to thermally populate magnetic excitations. At these temperatures the susceptibility of our samples is practically identical to the best quality samples; the results of the neutron scattering study we report and their interpretation are therefore unlikely to depend on sample quality. 

\begin{figure}
\includegraphics[width=8.25cm]{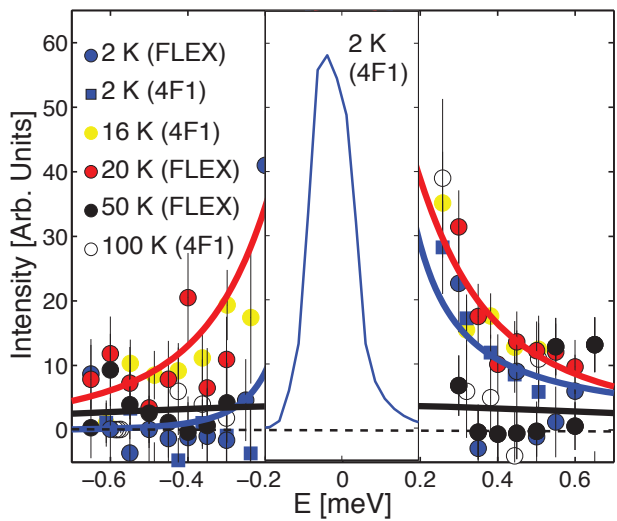}
\caption{
Constant ${\bf{Q}}$ = (200)  scans at several temperatures for FLEX (7 g with E$_{f}$=5.0 meV) and 4F1 (5 g with E$_{f}$=4.5 meV). The background scattering determined from rotating the sample away from Q=(200) has been subtracted. The inset shows the elastic peak divided by 150. The solid curves are fits to 20 K (red), 2 K (blue) and 100 K (black) to the equations described in the text.}
\label{fig1}
\end{figure}

Neutron inelastic scattering measurements were performed with the FLEX, 4F1 (LLB), and the MACS (NIST) \cite{Rodriguez08:19} cold triple axis spectrometers.   Beryllium filters were used on all instruments to filter out higher order neutrons.  Graphite PG(002) was used as both a monochromator and analyser with 4F1 and FLEX  (vertically focussed monochromators) and MACS (vertical and horizontal focussing).  The samples were aligned in the (HK0) scattering plane for the FLEX and 4F1 experiments and in the (0KL) plane for MACS.

The magnetic neutron scattering intensity at vector ${\bf Q}$ is directly related to the dynamic susceptibility by  
\begin{equation}
 I({\bf{Q,\omega}})
 \propto [n(\omega)+1]  {{\mid{F_{\bf Q}}\mid}^2}
 \sum_{\alpha,\beta}(\delta_{\alpha,\beta}-\hat{Q}_{\alpha}\hat{Q}_{\beta})
\operatorname{Im} {\chi_{\alpha \beta}(\bf{q},\omega)} 
\label{e0}
\end{equation}
The factor $F_{\bf Q}$ accounts for both the magnetic form factor and structure factor originating from uranium moments.  The $\alpha,\beta$ indices describe the moment directions before and after scattering and $n(\omega)$ is the Bose factor. Taking the axes along the principle axes of the orthorhombic structure the susceptibility tensor $\chi_{\alpha \beta}\equiv \delta_{\alpha\beta} \chi_\alpha$ is diagonal but not necessarily isotropic (susceptibilities are in $cgs$ units throughout this letter). For a nearly ordered itinerant ferromagnet at low energy and low momentum $\operatorname{Im} \chi_\alpha(\bf{q},\omega)$ for  $T>T_\text{Curie}$ is dominated by overdamped modes parameterised by \cite{lon99}
\begin{equation}
   \frac{\operatorname{Im} \chi_{\alpha}(\bf{q},\omega)}{\omega}= \chi_{\alpha}({\bf q})
   \frac{\Gamma_{\bf q}^\alpha}{\omega^{2}+{\Gamma_{\bf q}^\alpha}^{2}}
\label{e2}
\end{equation}
with $\Gamma_{\bf q}^\alpha = \gamma_q^\alpha / \chi_\alpha({\bf q})$ and $\chi_\alpha(q) \equiv  \chi_ \alpha(q,\omega=0)$. $I({\bf{Q}}, \omega)$ is therefore proportional to the Lorentzian in Eqn. \ref{e2}. NMR \cite{iha10} and susceptibility measurements \cite{huy08} are suggestive that $\chi_c$ dominates in the neutron cross-section.

\begin{figure}
\includegraphics[width=7.0cm]{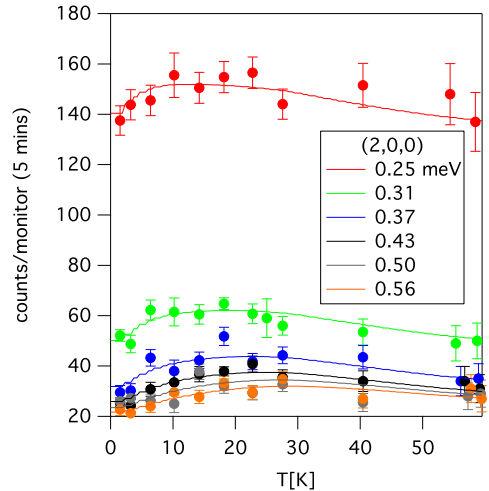}
\caption{
Neutron scattering counts/monitor for different temperatures and energy at ${\bf q}=(200)$ measured on 4F1. The solid lines are for a simultaneous fit to all the data with a single free parameter $\gamma_0 = 0.41(3) \mu eV$ (plus a temperature independent background term) as described in the text. }
\label{fig2}
\end{figure}

We first discuss our measurements in the HK0 scattering plane that characterise the energy scale of the dynamics.  Representative constant-$\bf{Q}$=(200) scans are illustrated in Fig. \ref{fig1} and demonstrate the presence of magnetic fluctuations which increase from almost zero intensity as the temperature is lowered from 100 K to 20 K.  The full temperature dependence for different energy transfers is shown in Fig. \ref{fig2}.  The intensity is peaked at a temperature which increases with increasing energy transfer.  The resolution is sufficiently narrow ($<0.1 \text{~meV}$) that contamination from the elastic peak can be neglected.

Since we are measuring at the zone centre, $\chi_c({\bf q}=0)$ can be replaced by the measured uniform susceptibility $\chi_c$ (determined with an MPMS magnetometer). Separate measurements of the elastic peak intensity with polarised neutrons (4F1) confirmed that the small sample studied with the MPMS is indeed representative of the whole crystal. Apart from the temperature independent background scattering at each energy and an overall constant of proportionality, the only undetermined parameter required to fit the entire data set at all temperatures and energies is $\gamma_{{\bf q}=0} \equiv \gamma_0$; on both theoretical grounds and by comparison with UGe$_2$ $\gamma_0$ is temperature independent.  A fit to the data gives $ \gamma_{0} = 0.41(3) \mu eV$ a value that is comparable to $\gamma_0[\text{UGe$_2$}]=0.7 \mu eV$ at room pressure. Landau damping vanishes as $q \rightarrow 0$. Its magnitude can be estimated from the fermi velocity (derived from the slope of the upper critical field at $T_s$) and the electronic density of states (derived from the heat capacity) to be $ \gamma_{q} \approx 2.5 [\mu\text{eV}/\text{\AA}^{-1}] \times q [\text{\AA}^{-1}]$ similar to that of ZrZn$_2$ \cite{ber88}. For UCoGe the observed damping is much larger than this for all $q$ within the resolution volume probed. Non-Landau damping must therefore dominate the magnetic dynamics in UCoGe as found for UGe$_2$. The observed peak in scattering as a function of temperature at each $\omega$ can be understood analytically for a Curie law susceptibility, noting this gives a peak in scattering at the temperature at which $\chi_\text{c}(T) \approx \gamma_0/ \omega$. 

We next examine how the magnetic excitations depend on wave vector moving away from the zone centre. The scattering can be modelled with 
\begin{equation}
\chi({\bf q})=\frac{1}{\chi_{0}^{-1}+(\xi_0^a q_a)^{2}+(\xi_0^b q_b)^{2}+(\xi_0^c q_c)^{2}} 
\label{e4}
\end{equation}
and $\Gamma({\bf q})=\gamma_q / \chi({\bf q}) \approx \gamma_0 / \chi({\bf q})$. For the self-consistent mode coupling theory the temperature dependence is completely derived from $\chi_0$. The $\xi_0^{\alpha}$ are temperature independent while the actual correlation lengths $ \xi_\alpha^2 = \chi_0 (\xi_0^\alpha)^2 $ ($\alpha=a,b,c$) diverge at $T_\text{Curie}$.  $\xi_0$ (and $\gamma_0$) are the key parameters deduced from our measurements and are not expected to be sample dependent; changes of $T_\text{Curie}$ would only be expected to change $\chi_0(T)$. 

\begin{figure}
\includegraphics[width=6.5cm]{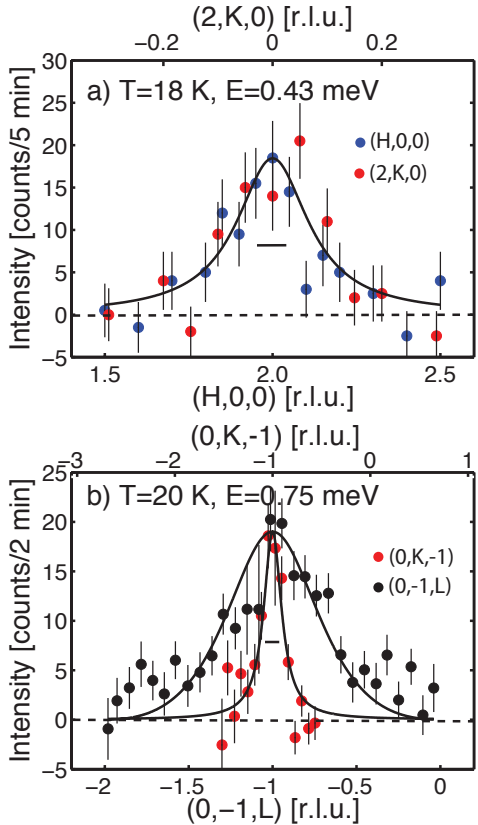}
\caption{
Constant energy scans taken on $a$) 4F1 and $b$) MACS.  The data on MACS ($b$) has been integrated over K=[-1.1,-0.9] and L=[-1.15,-8.5] and taken with E$_{f}$=5.0 meV with the samples (7 g+14 g) aligned in the (0,K,L) scattering plane.  The data taken on 4F1 is for 18 K and with E$_{f}$=4.5 meV with the sample (5 g) in the (H,K,0) scattering plane. The horizontal scales equate reciprocal wavelengths between the different axes.}
\label{fig3}
\end{figure}

\begin{figure}
\includegraphics[width=6.5cm]{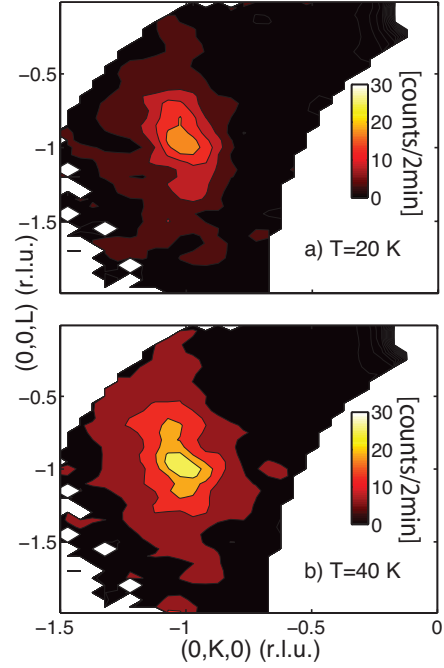}
\caption{
Constant E=0.75 meV slices taken using the MACS at T=20 K ($a$) and T=40 K ($b$) done near $\vec{Q}$=(0,-1,-1).  A background at T=1.5 K has been subtracted from both data sets. The two samples (7 g + 14 g) were aligned in the (0KL) scattering plane.}
\label{fig4} 
\end{figure}

Figure \ref{fig3} compares constant $\omega$ scans as $q$ moves away from the zone centre along different directions. The correlation lengths along the ${\bf a}$ (H) and ${\bf b}$ (K) directions are almost equal. The value $\xi_0^{a,b}= 86(21) \text{\AA}$  can be compared to the value of $\xi_{0\text{~UGe$_2$}}=152 \text{\AA}$. The scattering is however significantly broader along the {\bf c}-direction (L), which is the easy magnetic axis. The anisotropy is illustrated in figure \ref{fig4} which shows constant energy contours at 20 and 40 K for the whole ${\bf bc}$ plane.  The anisotropy in the (0KL) scattering plane at 20 K is  $\xi_c/\xi_{a,b} = 0.4$ based on fits to a lorentzian lineshape; the value of  $\xi_0^c = 34$ \AA.  The correlation length is thus significantly shorter in the direction parallel to the moment than perpendicular to it. In contrast, for UGe$_2$ the correlation length is almost isotropic \cite{hux03a}.  

We now move on to discuss the implications of our findings for SC  beginning with the energy scale of the magnetic excitations. $\Gamma(0) = \gamma_0/\chi$ has a strong pressure and temperature dependence given by $\chi$.  For UCoGe a conservative estimate of  the minimum value of $\chi$ in the region supporting SC is $\chi>5\times10^{-3}$.  This gives an energy scale $\Gamma(0) < 0.8 \text{~meV} \approx \text{~8 K}$ that is only just over an order of magnitude higher than $T_s$; this energy scale appears as the pre-factor in expressions for $T_s$ analogous to the Debye energy in BCS theory \cite{fay80}. The relatively modest ratio by which it exceeds $T_s$ is consistent with magnetic fluctuations driving superconductivity but implies that the coupling strength must be strong.  

We now discuss how the magnetic lengthscales we find point to an important role for small Fermi-surface pockets in the superconducting mechanism.  For a single isotropic parabolic band $\xi_0^2 = 1/(12 \chi_p k_f^2)$ with $\chi_p \equiv \mu_B^2 N(0)$ the non-exchange enhanced Pauli susceptibility \cite{lon85} and $N(0)$ the electronic density of states (for both spins). For an anisotropic effective mass $\xi_{0}^{x2}=[(m_x m_y m_z)^{1/3}/m_x] \xi_0^2$ with $k_f \equiv (3 \pi^2 n)^{1/3}$ and $n$ the electron density. 
The inverse correlation length determines the range of Fermi-wavevectors that are mixed by magnetic interactions.  For odd-parity SC the order parameter changes sign with direction. An inverse correlation length longer than $2 k_f$ is therefore detrimental to SC, suppressing it for $k_f^2 \xi^2 = \chi_0/(12 \chi_p)<<1$. The inequality is not satisfied and the suppression avoided if $\chi_0$ is enhanced relative to $\chi_p$ as found close to a ferromagnetic quantum critical point. In the opposite limit $k_f\xi >>1$ the pairing interaction is effective over an increasingly restrictive fraction of the Fermi-surface and $T_s$ is again reduced. $T_s$ is a maximum for $k_f \xi \approx 1$ as may be anticipated combining these two limits; the expression describing this appears in the exponential of the  BCS like expression for $T_s$ \cite{fay80}.  
The Fermi surface in UCoGe has been calculated to comprise both small pockets and large surfaces spanning the Brillouin zone\cite{sam10}. For $k_f \approx 0.5 \text{\AA}^{-1}$, spanning the Brillouin zone,  $k_f \xi>>1$ at low temperature with the optimal condition $k_f \xi \approx 1$ only reached at around 40 K, well above $T_s$. At low temperature $k_f \xi \approx 1$ can however be satisfied for small Fermi-surface pockets as observed to exist in UCoGe by quantum oscillation measurements~\cite{aok11}. 

The anisotropy of the correlation length infers that the Fermi-velocity is isotropic in the ${\bf ab}$ plane but 2.4 times smaller along ${\bf c}$. The anisotropy should directly carry over to the slope of the critical field for SC at $T_s$ (corrections owing to the anisotropic order parameter are small). H$_{c2}$ close to $T_s$ is isotropic in the ${\bf ab}$ plane and smaller along ${\bf c}$ \cite{huy07} \cite{aok09}, but by a factor of $10$ rather than $2.4$. Our finding that small Fermi-surface pockets may be advantageous for superconductivity, combined with the calculated sensitivity of the size of these pockets to magnetic polarisation \cite{sam10} suggest that changes in pocket size may strongly suppress $H_{c2}^{\bf c}$. This could account for the enhanced anisotropy of the critical field close to $T_s$, rather than a strong field dependent enhancement of the pairing for fields perpendicular to the ${\bf c}$ axis; the latter explanation was suggested in Ref. \onlinecite{aok09} to account for the unusual critical field dependences at lower temperature.  

The anisotropy we have measured also has implications for the symmetry of the superconducting order parameter. SC is enhanced by increasing the Fermi-surface integral of the product of $\chi m_B/m$ weighted by the gap function where $m/m_B$ is the electronic mass enhancement ($m_B$ is the band mass).  $\chi$ reflects the response of the entire Fermi-surface and anisotropy comes from the mass enhancement. Based on this, gap nodes would be located preferentially along the directions of maximum mass enhancement. For UCoGe, assuming the mass enhancement mirrors the effective mass anisotropy deduced from the anisotropic correlation length this would favour a gap symmetry with nodes located along the $c$-axis\cite{min04}.  

While a calculation of $T_s$ for magnetically mediated SC requires a detailed knowledge of the Fermi-surface, which is currently not available, the qualitative analysis discussed above suggests that the magnitudes of the correlation lengths and their anisotropy have traits that could promote magnetically mediated SC, potentially with the largest gap occurring on small Fermi-surface pockets and gap nodes parallel to ${\bf c}$. 
 
\noindent {\it acknowledgements}  
Support from the Royal Society (AH), EPSRC (AH, DS) and SUPA (CS) is gratefully acknowledged. Work at Los Alamos National Laboratory was performed under the auspices of the U.S. DOE, OBES, Division of Materials Sciences
and Engineering and funded in part by the LANL LDRD program.
\end{document}